\documentclass{PoS}

\usepackage{epsfig}
\usepackage{graphicx}

\title{Nucleon-Nucleon Chiral Two Pion Exchange potential vs
Coarse grained interactions}

\ShortTitle{NN Two Pion Exchange}

\author{\speaker{Rodrigo Navarro Perez}%
         \thanks{Work supported by Spanish DGI
  (grant FIS2011-24149) and Junta de Andaluc{\'{\i}a} (grant FQM225).
  R.N.P. is supported by a Mexican CONACYT grant.}\\
Departamento de F\'{\i}sica At\'{o}mica, Molecular y 
  Nuclear and Instituto Carlos I de \\ F{\'\i}sica Te\'orica y Computacional. 
Universidad de Granada, E-18071 Granada, Spain.\\
        E-mail: \email{rnavarrop@ugr.es}}

\author{J. E. Amaro\\
Departamento de F\'{\i}sica At\'{o}mica, Molecular y 
  Nuclear and Instituto Carlos I de \\ F{\'\i}sica Te\'orica y Computacional. 
Universidad de Granada, E-18071 Granada, Spain.\\
        E-mail: \email{amaro@ugr.es}}

\author{E. Ruiz Arriola\\
Departamento de F\'{\i}sica At\'{o}mica, Molecular y 
  Nuclear and Instituto Carlos I de \\ F{\'\i}sica Te\'orica y Computacional. 
Universidad de Granada, E-18071 Granada, Spain.\\
        E-mail: \email{earriola@ugr.es}}

\abstract{ We analyse the interplay between nucleon-nucleon potentials
  deduced from chiral perturbation theory and a coarse grained
  representation of the short distance interactions by delta-shells
  potentials below a certain cut-off distance. While we find that
  the number of parameters is greatly reduced when Chiral Two Pion
  Exchange contributions are included we also observe that discerning the
  necessity of improvements on the interaction requires a detailed
  analysis of all error sources. Our points are best illustrated by
  computing deuteron static properties as well as electromagnetic form
  factors after error propagation.}

\FullConference{The 7th International Workshop on Chiral Dynamics,\\
		August 6 -10, 2012\\
		Jefferson Lab, Newport News, Virginia, USA}
\begin{document}

\section{Introduction}

The chiral theory of Nuclear Forces has become a popular approach in
recent years~\cite{Epelbaum:2008ga,Machleidt:2011zz}. Indeed, while
Charge Dependence and One Pion Exchange provided a satisfactory fit to
np and pp data~\cite{Stoks:1993tb} leading to high quality potentials
used for Nuclear
applications~\cite{Stoks:1993tb,Wiringa:1994wb,Machleidt:2000ge,Gross:2008ps},
Chiral Two Pion Exchange potentials~\cite{Kaiser:1997mw} have improved
the analysis~\cite{Rentmeester:1999vw,Rentmeester:2003mf}.

In this contribution we reanalyze the problem directly in terms of a
delta-shell potential which gives a simple way to coarse grain the
interaction between two nucleons down to the relevant shortest de
Broglie wavelength~\cite{NavarroPerez:2011fm}. This form was first
introduced by Aviles \cite{Aviles:1973ee} and has recently been used
to calculate nuclear binding energies~\cite{NavarroPerez:2011fm}, to
extract~\cite{NavarroPerez:2012qf} and
propagate~\cite{NavarroPerez:2012vr,Perez:2012kt} the corresponding
uncertainties inherent to the NN interaction or to evaluate the
effective interactions~\cite{NavarroPerez:2012qr}.

\section{Delta-shell and Chiral Potentials}

In our analysis the potential consists of a short range piece and a
long range contribution as follows
 \begin{equation}
   V(r) = \sum_{n=1}^{18} O_n \left[\sum_{i} V_{i,n} r_i \delta(r-r_i) \right] + \Big[ V_{\rm OPE}(r) + V_{\rm TPE1o}(r) + V_{\rm  TPEso}(r) + V_{\rm em}(r) \Big] \theta(r-r_c),
\label{eq:potential}
\end{equation}
where $O_n$ are the set of operators in the
AV18 basis~\cite{Wiringa:1994wb}, $r_i$ are the concentration radii and
$V_{i,n}$ are strength coefficients, which are used as fitting
parameters.  For definiteness $V_{\rm OPE}(r) $, $ V_{\rm
  TPElo}(r)$, $ V_{\rm TPEso}(r) $ and $V_{\rm em}(r)$ are
those of Ref.~\cite{Rentmeester:1999vw}. The distance between the
delta-shells $\Delta r$ is determined from the shortest de Broglie
wavelength (for a detailed explanation see the appendix in
\cite{Entem:2007jg} and \cite{NavarroPerez:2011fm}) below pion
production threshold i.e. $\Delta r = 1/\sqrt{M_N m_\pi}  \sim 0.6$fm, so that $r_i = i \Delta
r \le r_c$. Our purpose is to see how small can $r_c$
become when 
$V_{\rm OPE}(r) $, $ V_{\rm TPElo}(r)$ and $ V_{\rm TPEso}(r) $ in Eq.~(\ref{eq:potential}) are subsequently
added.  

%The range of validity of the long distance interaction can be
%probed by varying $r_c$. For every potential we used three different
%values for $r_c$ $1.8$fm, $2.4$fm and $3.0$fm.

\section{Chiral TPE vs OPE}

\begin{figure}[ht]
\begin{center}
\epsfig{figure=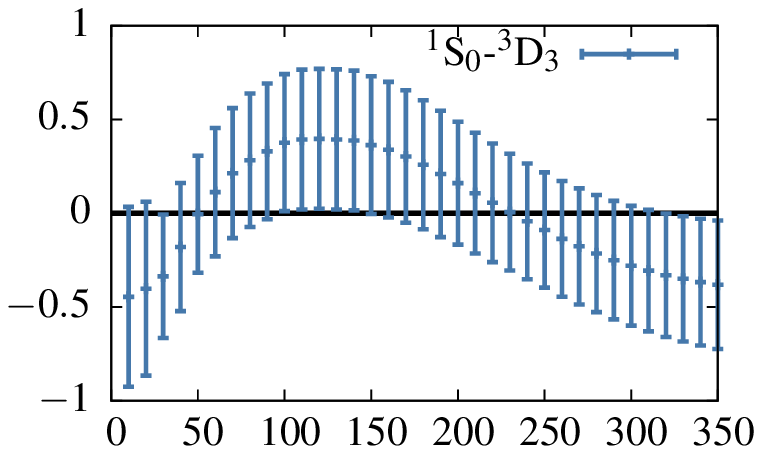,height=3cm,width=4.9cm}
\epsfig{figure=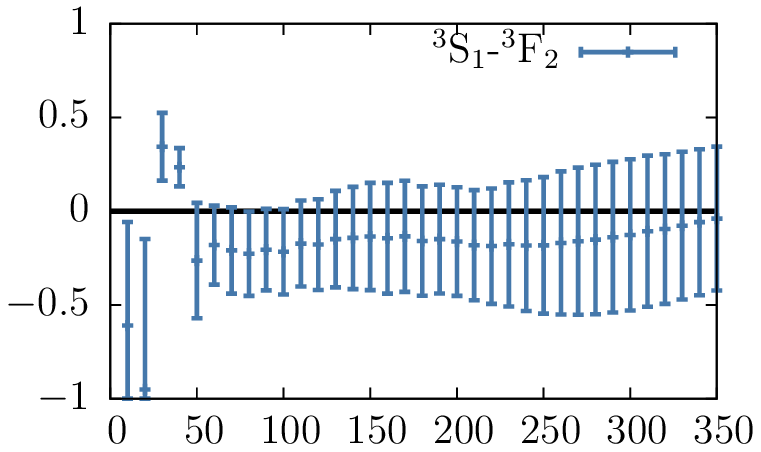,height=3cm,width=4.9cm}
\epsfig{figure=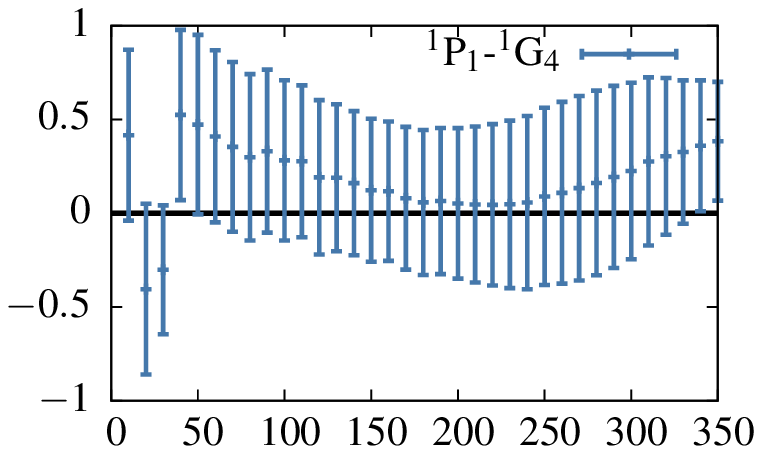,height=3cm,width=4.9cm}
\end{center}
\caption{Correlations among different phase-shifts of the PWA and six
  high quality
  potentials~\cite{Stoks:1993tb,Wiringa:1994wb,Machleidt:2000ge,Gross:2008ps}
  which provided a $\chi^2/ {d.o.f} \lesssim 1 $. 
The correlation factor was calculated using the equation
$    r_{x,y} = \sum_{i=1}^{n}{(x_i-\bar{x})(y_i-\bar{y})}/(n \sigma_x \sigma_y)
$ where the bar indicates the mean of a variable and $\mathbf{\sigma}$ the
corresponding standard deviation.}
\label{FigCorrelations}
\end{figure}

As a preliminary step in our analysis we fitted
Eq.~(\ref{eq:potential}) to a pseudo-database constructed from the np
phase-shifts given by the 1993 Partial Wave Analysis and the
subsequent phase-shifts of the 6 high quality potentials that give a
$\chi^2/\nu \lesssim 1$ when compared to experimental scattering
data~~\cite{Stoks:1993tb,Wiringa:1994wb,Machleidt:2000ge,Gross:2008ps}.
Given this, we have two alternatives based on the treatment of this
pseudodata. Either we make accurate fits to {\it each} single
potential phase-shifts and we average the seven different results and
determine their mean squared deviation or we asign a mean value and an
error to the compilation as a whole and make a standard fit taking the
pseudodata with errors as experimental data. While the first procedure
seems to be a quite natural way to incorporate correlations between
different partial waves, it turns out that these correlations are
almost negligible as can be seen for some representative cases in
Fig.~(\ref{FigCorrelations}).

\begin{figure}[p]
\begin{center}
\epsfig{figure=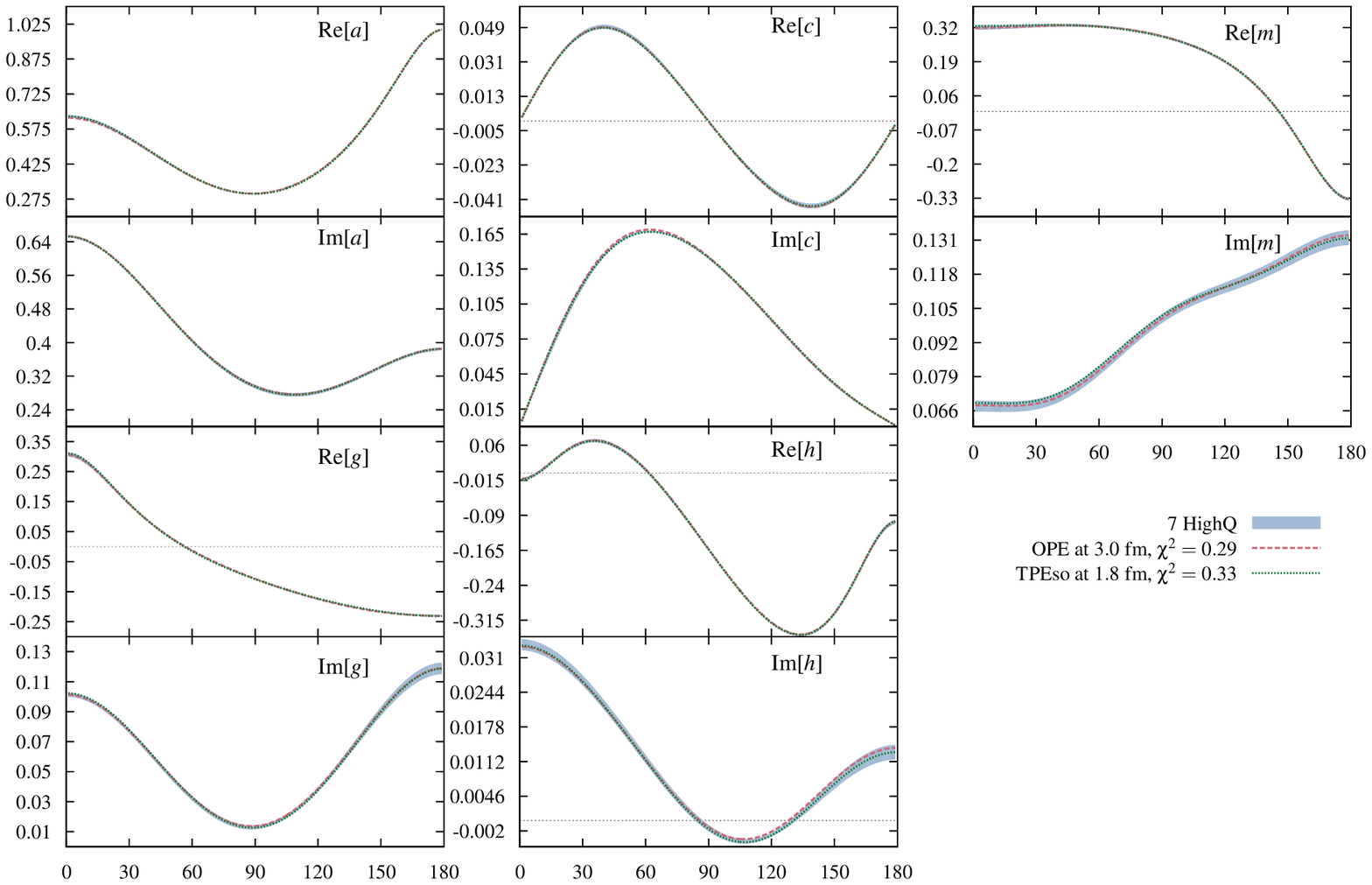,height=9cm,width=14.7cm}
\epsfig{figure=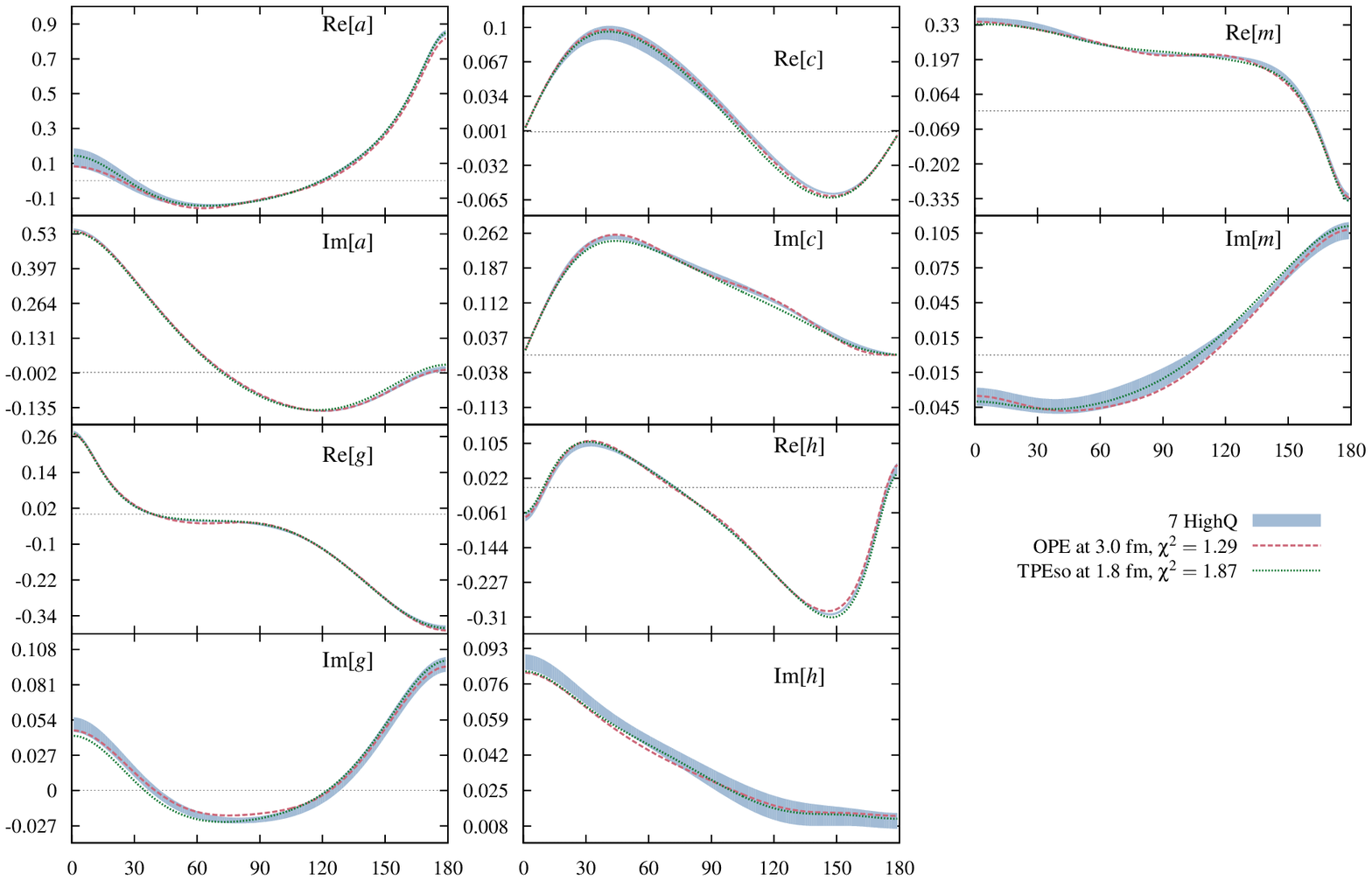,height=9cm,width=14.7cm}
\end{center}
\caption{np Wolfenstein parameters for different energies in the
  laboratory system as a function of the CM angle. Upper panel:
  $E_{\rm LAB}=100 {\rm MeV}$. Lower panel:
  $E_{\rm LAB}=350 {\rm MeV}$. The band represents the compilation
  of the PWA and six high quality
  potentials~\cite{Stoks:1993tb,Wiringa:1994wb,Machleidt:2000ge,Gross:2008ps}
  which provided a $\chi^2/ {d.o.f} \lesssim 1 $. The
  dashed line denotes the results obtained by the fitted interaction with
  OPE and $r_c = 3.0$fm, while the doted line comes from the interaction
  with TPE and $r_c = 1.8$fm.}
\label{FigWolfenstein}
\end{figure}

Table~(\ref{TableFitsPhaseShifts}) shows the value of $\chi^2/\nu$ and the
number of parameters for every potential depending on the long range interaction
and the radial cut-off. The results show, in agreement with previous findings,
that for OPE and TPElo the range of validity is between $1.8$fm and $2.4$fm since using
$r_c= 1.8$fm no longer gives a satisfactory fit. With TPEso one can go down to $r_c = 1.8$fm
and even reduce the number of parameters needed.
\begin{table}
\centering
\begin{tabular}{c c c c c c c}
 $r_{\rm c}$ [fm] & 1.8 &               & 2.4 &               & 3.0 &                \\
\hline  
                  & \#p & $\chi^2/\nu$  & \#p & $\chi^2/\nu$  & \#p & $\chi^2/\nu $ \\
 \hline\noalign{\smallskip}
            OPE   & 37  &   2.1383         & 47  &   0.6470      & 51  &   0.4653      \\ 
            TPElo & 40  &   2.0661         & 46  &   0.7361      & 52  &   0.5047      \\ 
            TPEso & 32  &   0.5911         & 44  &   0.5225      & 51  &   0.3928                  
\end{tabular}
\caption{$\chi^2/\nu$ and number of parameters for fits to phasehifts}
\label{TableFitsPhaseShifts}
\end{table}

Just as the PWA and the 6 high quality potential show a dispersion on
the phaseshifts the same occurs with the scattering amplitude. This
can be easily seen by using the Wolfenstein decomposition of the
scattering amplitude~\cite{Golak:2010wz},  
\begin{eqnarray}
   M(\mathbf{k}_f,\mathbf{k}_i) &=& a(\theta,p) + m (\theta,p) (\mathbf{\sigma}_1,\mathbf{n})(\mathbf{\sigma}_2,\mathbf{n}) 
                  + (g(\theta,p)-h(\theta,p))(\mathbf{\sigma}_1,\mathbf{m})(\mathbf{\sigma}_2,\mathbf{m}) \nonumber \\
                  & &+ (g(\theta,p)+h(\theta,p))(\mathbf{\sigma}_1,\mathbf{l})(\mathbf{\sigma}_2,\mathbf{l})  + c(\theta,p)(\mathbf{\sigma}_1+\mathbf{\sigma}_2,n) \, , 
\end{eqnarray}
and comparing the 5-complex Wolfenstein parameters for every
interaction.  Since all the scattering observables can be directly
calculated from the Wolfenstein parameters the dispersion on the
amplitude can be a measure of the dispersion on observables as
well. With this in mind we calculated the Wolfenstein parameters of
the potentials in table~(\ref{TableFitsPhaseShifts}) as a function of
laboratory energy $T_{\rm LAB}$ and scattering angle $\theta$ and
compared them to the mean of the PWA and 6 high quality potentials
using the standard deviation as the uncertainty to calculate a
$\chi^2/\nu$.  The results are shown in
table~(\ref{TableFitsWolfenstein})
\begin{table}
\centering
\begin{tabular}{c c c c c c c}
 $r_{\rm c}$ [fm] & 1.8           & 2.4           & 3.0           \\
                  & $\chi^2/\nu$  & $\chi^2/\nu$  & $\chi^2/\nu $ \\
            \hline\noalign{\smallskip}
            OPE   &   2.45        &   0.56        &   0.47        \\ 
            TPElo &   2.92        &   0.69        &   0.49        \\ 
            TPEso &   0.54        &   0.70        &   0.41             
\end{tabular}
\caption{$\chi^2/\nu$ to wolfenstein parameters}
\label{TableFitsWolfenstein}
\end{table}
and exhibit very similar features to the ones in
table~(\ref{TableFitsPhaseShifts}).  Figure~(\ref{FigWolfenstein})
shows the disperssion of the Wolfenstein parameters for the high
quality potentials as a function of the scattering angle $\theta$ at
$T_{\rm LAB}=100$MeV and $T_{\rm LAB}=350$MeV.  We also show the
Wolfenstein parameters given by the coarse grained interactions with
OPE using $r_c = 3.0$fm and TPEso with $r_c = 1.8$fm.

\section{Deuteron Properties}

For a comparison between OPE and (chiral) TPE we calculate a few deuteron
properties with the potentials constructed in this contribution. The results
are shown in table~(\ref{TableDeuteronProperties}) and show no significant
diference on the central values and very similar uncertainties, being all of
them compatible with previously known empirical values, and reflecting the pseudodata uncertainties. 
\begin{table}
\centering
\begin{tabular}{cccccccc}
 Potential & $r_c $(fm) & $B_D$(MeV)  &   $\eta$   &  $A_S$ (fm$^{1/2}$)    & $P_D$    &  $r_m$(fm)    &   $ Q_D$ (fm$^2$)     \\
%&   $\langle r^{-1} \rangle$  (fm$^{-1}$)\\
%       & fm   & MeV    &            &fm$^{1/2}$  &  \%      &  fm       &   fm$^2$   &   fm$^{-1}$ \\
 \hline\noalign{\smallskip}
 OPE  &  $3.0$    &-2.2(2) &   0.025(2) &   0.88(3)  &   5.7(2) &   1.97(8) &   0.272(9) \\
%&   0.45(1) \\
% TPElo  &  $3.0$   &-2.2(2) &   0.025(2) &   0.88(3)  &   5.7(2) &   1.97(8) &   0.272(9) &   0.45(1) \\
% TPEso   &  $3.0$  &-2.2(2) &   0.025(2) &   0.88(3)  &   5.7(2) &   1.97(8) &   0.272(9) &   0.45(1) \\
% \hline\noalign{\smallskip}
% OPE  &  $2.4$    &-2.2(3) &   0.025(2) &   0.89(4)  &   5.6(3) &   2.0(1)  &   0.27(1)  &   0.45(1) \\
 TPElo  &  $2.4$  &-2.2(3) &   0.025(2) &   0.89(4)  &   5.6(3) &   2.0(1)  &   0.27(1)  \\
% &   0.45(1) \\
% TPEso  &  $2.4$  &-2.2(5) &   0.025(4) &   0.89(8)  &   5.5(6) &   2.0(2)  &   0.27(2)  &   0.46(3) \\
% \hline\noalign{\smallskip}
% OPE   &  $1.8$   &-2.2(3) &   0.025(3) &   0.88(5)  &   5.7(4) &   2.0(1)  &   0.27(1)  &   0.46(2) \\
% TPElo &  $1.8$   &-2.2(6) &   0.025(5) &   0.88(9)  &   5.7(6) &   2.0(2)  &   0.27(2)  &   0.46(3) \\
 TPEso  &  $1.8$  &-2.2(4) &   0.025(3) &   0.88(6)  &   5.6(4) &   2.0(1)  &   0.27(2)  \\
%&   0.45(2) \\      
 \hline\noalign{\smallskip}
 Empirical& & -2.2245(2) & 0.0256(5)  & 0.8781(44) &  5.67(4) &  1.953(3) & 0.2859(3)         
\end{tabular} 
\caption{Deuteron properties. Notation is as follows, $B_D$ binding energy,
$\eta$ assymptotic ratio, $A_S$ S-state normalization, $P_D$ D-state probability,
$r_m$ root mean square radius and $Q_D$ electric quadrupole moment.}
\label{TableDeuteronProperties}
\end{table}

\begin{figure}[pb]
\begin{center}
\epsfig{figure=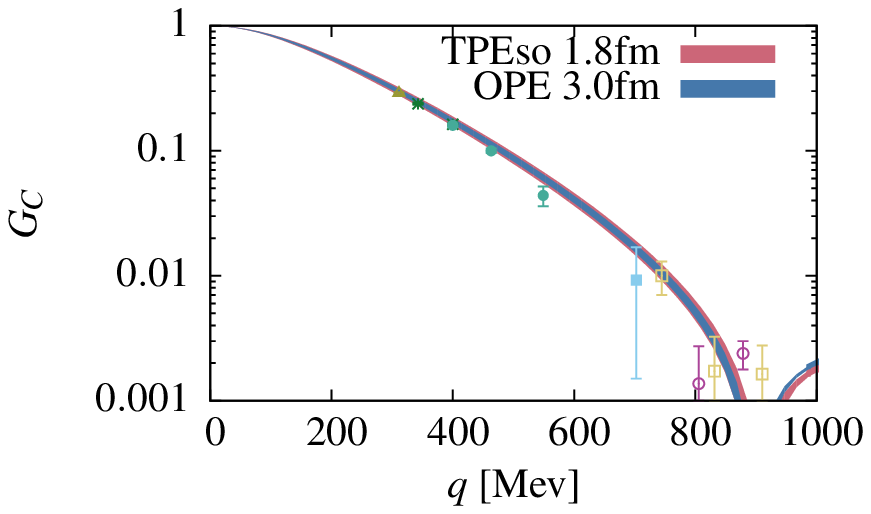,height=4cm,width=4.9cm}
\epsfig{figure=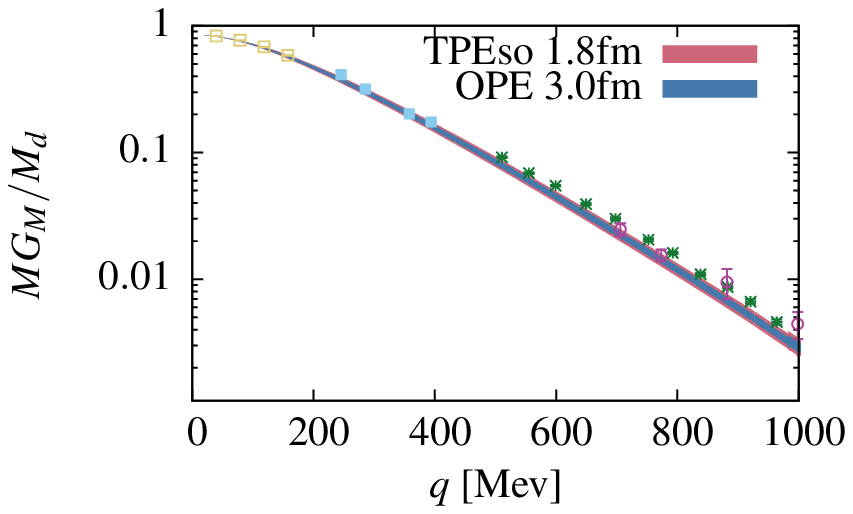,height=4cm,width=4.9cm}
\epsfig{figure=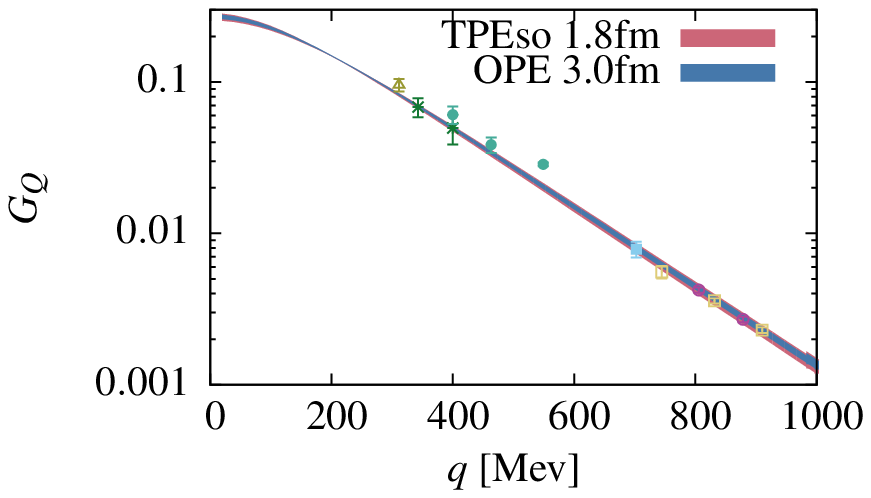,height=4cm,width=4.9cm}
\end{center}
\caption{Deuteron Form Factors from OPE with $r_c = 3.0$fm (blue band)
and TPE with $r_c = 1.8$fm (red band). The error bar was obtained by
propagating the uncertainty from the pseudodata as explained in the text.} 
\label{FigFormFactors}
\end{figure}

Deuteron form factors using OPE with $r_c=3.0$fm and TPElo with
$r_c=1.8$fm are presented Fig.~(\ref{FigFormFactors}) with propagated
uncertainties. As we see there is no significant difference between
using OPE or TPE as the long range np interaction. The rather small
discrepancy between calculated and experimental values could be
resolved by the inclusion of Meson Exchange Currents (MEC).  In the
$G_C$ form factor we see that within errors there is no discrepancy.

\section{Conclusions}

In the present contribution we have adressed a comparison between the
well-known OPE potential and the chiral TPE interactions. The short
distance piece of the potential is represented by a delta-shells
potential which features a coarse graining of the unknown physics down
to the smallest de Broglie wavelength probed by the NN interactions
below pion production threshold. The long range part is assumed to be
valid down to a radial cut-off distance $r_c$ and we analyze the
quality of each fit as a function of this distance. For our analysis
we use pseudodata consisting of a compilation of the np phase shifts
given by 7 high quality fits~\cite{Stoks:1993tb,Wiringa:1994wb,Machleidt:2000ge,Gross:2008ps}.
The error asignment corresponds to a lack of correlation between
different partial waves; a circumstance which turns out to be true
within the inherent uncertainties of the different potentials. There
is substantial reduction in the number of parameters needed for the
short range part of the interaction.  Indeed for OPE, one has $r_c =
3.0$ fm, $\chi^2/\nu = 0.47$ and $51$ parameters are needed whilst OPE
+ (chiral) TPEso, gives $r_c = 1.8$ fm, $\chi^2/\nu = 0.59$ and the number
of parameters is reduced to $32$. From a Nuclear Physics Structure
point of view it is uncertain what could be the real advantage in
implementing as a matter of principle the chiral TPE
interaction. Actually, to decide objectively on this issue requires a
meticulous determination of both statistical and systematic errors. We
have illustrated this point by computing the deuteron form factors and
propagating the corresponding uncertainties deduced by the error
treatment of the pseudodata. This is a crucial issue to discern on the
real role of the MEC conributions to the form factors. For instance,
the charge form factor acquires purely transverse contributions which
have been estimated to be small. The question is whether or not the
size of the MEC's is larger than the estimated uncertainties.

%\begin{itemize}
% \item Sampling of the NN interaction by a delta shell potential
%% \pause
%% \item Sampling resolution determined by the deBroglie wavelength of the most energetic particle in our scheme
%% \pause
% \begin{equation}
%  1/\sqrt{m_\pi M}  \lesssim \Delta r \lesssim 1/m_\pi
% \end{equation}
% \item 3 well defined regions

% \item Fit to 
% \begin{itemize}
%  \item Every partial wave with $j \leq 5$
%  \item OPE, $r_c = 3.0$ fm, $\chi^2/\nu = 0.47$ and $51$ parameters
%  \item OPE + $\chi$TPE, $r_c = 1.8$ fm, $\chi^2/\nu = 0.59$ and $32$ parameters   
%%  \item Systematic error propagation
%%  \item Quantitative comparison of OPE and Chiral TPE
%%  \item Different $r_{\rm c}$ for the long range part of the interaction
%%  \item 40 fitting parameters
%%  \item $\chi^2/\text{d.o.f.} \lesssim 2$ (less than 1 in some waves)
% \end{itemize}
% \end{itemize}

%\bibliographystyle{abbrv}
%\bibliographystyle{unsrt}
%\bibliography{refs}

\end{document}